# Zn-indiffused diced ridge waveguides in MgO:PPLN generating 1 watt 780 nm SHG at 70% efficiency

SAM A. BERRY,[1,*] Lewis G. Carpenter,[1] Alan C. Gray,[1] Peter G. R. Smith,[1] AND Corin B. E. Gawith[1,2]

[1]*Optoelectronics Research Centre, Zepler Institute, University of Southampton, University Road, Southampton SO17 1BJ, United Kingdom*
[2]*Covesion Ltd., Premier Way, Romsey SO51 9DG, United Kingdom*
[*]*s.berry@soton.ac.uk*

**Abstract:** We present a metallic zinc indiffused diced ridge waveguide in magnesium doped periodically poled lithium niobate (MgO:PPLN) capable of generating over 1 W of 780 nm with 70% efficiency. Our 40 mm long waveguide has near circular fundamental mode output with diameter 10.4 µm and insertion loss of -1.17 dB. Using a commercial 2 W EDFA-based system, the SHG output power did not exhibit roll-off at maximum available pump power.

## 1. Introduction

Quantum technology (QT) is driving a renewed focus on nonlinear optical materials principally for converting the operating wavelength of established narrow linewidth lasers to an atomic cooling transition, and for generating entangled photon pair states. Waveguides in nonlinear materials, with their increased optical efficiencies at lower pump powers and small device footprint, have become an active area of research [1–4].

For atom traps, multiple laser frequencies are typically derived from a fundamental atomic cooling transition. To support large atom condensates, watt-level powers with linewidths below the hyper-fine atomic transition energies are desired [5]. Generating this laser output has been achieved using established narrow linewidth lasers and bulk periodically poled lithium niobate (PPLN) crystals [5–7], however these require >10 W pump sources limiting field-based applications in QT, such as portable gravity measurements. Nonlinear waveguide have been used to enable more efficient conversion of laser pump sources, with some commercial availability for <250 mW operation. With their high $\chi^{(2)}$ nonlinearities and wide transparency window, PPLN waveguides have been used for nonlinear interactions in telecommunications [8,9], quantum communications [10,11], single photon sources [11–14], all-fiber/waveguide squeezing [15], and atom/ion trapping [4,16–18]. In particular, the frequency doubling of 1560 nm sources is frequently chosen for Rb atom traps, which is the interest of our work.

Table 1 shows some examples of nonlinear optical waveguide technologies used to frequency double telecoms wavelengths, summarising the normalised conversion efficiencies and maximum output power achieved. For a fair comparison, this table was restricted to one wavelength of interest as waveguides operated in different ranges are not directly comparable; this is because of the efficiency dependence on frequency and other limitations such as photorefractive damage and green/blue induced infrared absorption (GRIIRA/BLIIRA) effects. The reported efficiencies have been renormalised with respect to pump power and with the maximum attained SHG power to enable a direct comparison between the different formats at their highest operating power. Whilst this figure of merit does not highlight the extremely high efficiencies that can be achieved in some waveguide geometries [20], we have removed the quadratic dependency on length for this table; the achievable device length is affected by the waveguide uniformity, which is fabrication dependent, and results in various device lengths often reported where the total SHG conversion is most efficient. Lack of waveguide uniformity can limit the overall efficiency of longer devices due to variation of the phasematching condition along its propagation length [8]; this makes fabricating longer devices that scale quadratically with length difficult using some manufacturing techniques.

**Table 1. PPLN waveguide CW SHG powers and efficiencies for the C-band**

| Waveguide type | $\eta_{norm}$ | $P_{max}$ |
|---|---|---|
| PPLN on $LiNbO_3$ [3] | 54.0 %/W | 1.7 W |
| PPLN on $LiTaO_3$ [19] | 2400 %/W | 150 mW |
| Nano- PPLN wvgs [20] | 416 %/W | 117 mW |
| Ti:PPLN ridge wvgs [18] | 10.35 %/W | 35 µW |
| Zn:PPLN channel wvgs [21] | 59 %/W | 5.2 µW |
| RPE PPLN waveguides [22] | 1633 %/W | NR |
| APE PPLN waveguides [11] | 63 %/W | NR |

Efficiencies have been recalculated to be normalised with power only. NR: Not reported.

While many of the devices listed in Table 1 show remarkable efficiency for milliwatt pump powers, watt-level powers of interest for atom trap gravimeters have previously only been reported in the process-intensive MgO:PPLN bonded to $LiNbO_3$ or $LiTaO_3$ format, where 5 W of 1550 nm pump was converted to 1.7 W at 775 nm for Rb atom cooling, representing 34% device efficiency [3]. A commercially available device has been shown, when pumped beyond its published specifications to 1.8 W of 1560 nm input power, to generate 1 watt of 780 nm representing a 56% conversion efficiency [16]. In this paper we address PPLN waveguides to achieve higher conversion efficiencies at operating powers above 1 watt.

## 2. Fabrication

Our PPLN waveguide manufacturing is based on three simplified steps of zinc indiffusion to create a planar guiding layer, ultra-precision machining to form single-mode ridge waveguides, and preparation of facets by dicing. We deposit a metallic zinc layer to create a uniform indiffusion for the planar guiding layer, ensuring a phasematching response close to the theoretical prediction and improvement over the use of ZnO used in our previous work [23], which describes the steps taken to optimise the indiffusion parameters and ridge widths for single-mode operation. Zinc indiffusion is performed post-poling to preserve the uniformity of the periodically poled domains; it would be possible to perform the zinc indiffusion before poling. Figure 1 illustrates the fabrication process: (1) a 100 nm thin film metallic zinc layer is sputtered onto a PPLN wafer (with poled period of 18.3 µm to facilitate 1560-780 nm SHG) using a commercial sputter tool; (2) this metallic layer is indiffused in a furnace with an oxygen atmosphere at around 950 °C, which is below the crystal's Curie temperature and ensures spontaneous domain inversion does not occur; (3) ridge waveguide structures are cut into the wafer using optical quality dicing to achieve sub-nanometer surface roughness [24–26]. The cut depth exceeds the indiffusion layer thickness (solid lines), ensuring the ridge waveguide geometry. Chip singulation is achieved with full-wafer thickness cuts (dashed lines). A resulting PPLN waveguide chip is depicted in Fig. 1(4), including multiple ridges. A scanning electron microscope (SEM) image of the waveguides after fabrication is shown in Fig. 2 where the optical quality of the dicing can clearly be seen with one cut defining the sides of two parallel ridge waveguides. For high power operation ridge waveguide geometries have been shown to reduce photorefractive damage in Ti-based diffused waveguides [18] and Zn indiffusion has been shown to provide high resistance to photorefractive damage in MgO:LiNbO$_3$ [27].

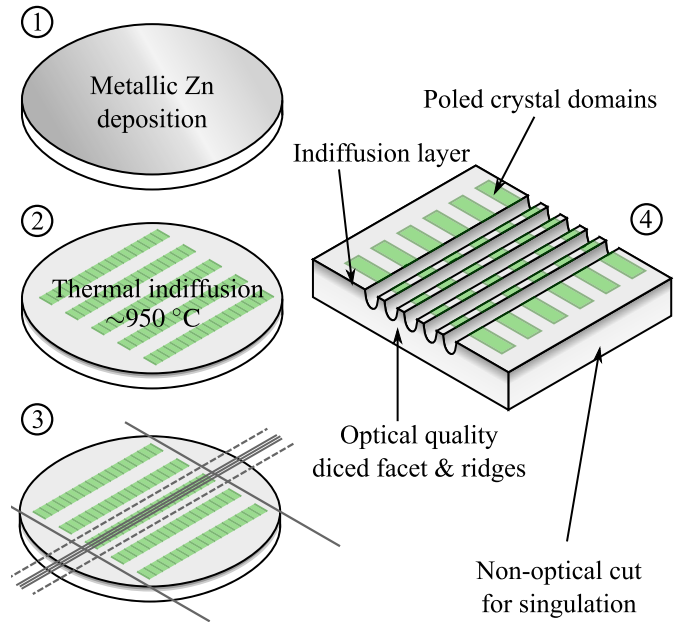

**Fig. 1.** The fabrication of the diced ridge PPLN waveguides: (1) a 100 nm metallic zinc layer is deposited onto a PPLN wafer, (2) the metallic zinc is indiffused below 950 °C in an O$_2$ atmosphere to create a planar guiding layer, (3) ridge waveguides are diced into the wafer (including facets) using ductile dicing, and (4) the resulting multiple ridge waveguide chip structure when singulated out of the wafer.

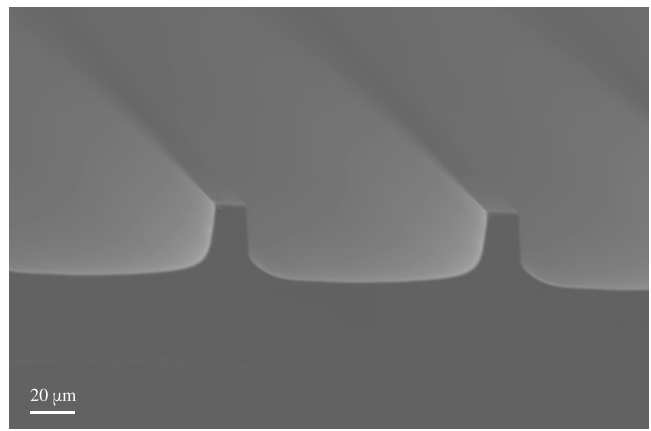

**Fig. 2.** Scanning electron microscope image of diced ridge waveguides in MgO:PPLN. Two ridge waveguides are shown here defined by three optical quality dicing cuts.

Lateral mode confinement affects the effective refractive index of the waveguide mode when close to the single mode condition. For SHG the ridge size is typically defined to be single-mode at the fundamental frequency, while at the SHG wavelength the waveguide is expected to be multimode (where width variations have a lesser effect on the effective refractive indices); thus any width variation along with waveguide will cause variation in the phasematching condition. Because an absolute waveguide width is difficult to achieve, our waveguide structure is designed to include multiple ridge widths between 11 µm and 13 µm to ensure the presence of a waveguide with the desired wavelength and temperature phasematching condition. The waveguides are single mode across the C-band, have a mode field diameter (MFD) well-matched to PM1550 fiber and have an insertion loss of -1.17 dB.

## 3. Experimental setup

Figure 3 details the experimental setup used to characterise our PPLN waveguides. A tunable laser source is used to sweep the pump wavelength around the central phase matching wavelength so that the operating temperature can be kept constant to avoid alignment issues caused by different thermal expansions in the optical setup. The tunable laser is amplified and is then passed through an adjustable collimator and aspheric lens to match the mode size of the PPLN waveguide. The crystal is mounted within a Covesion PV40 oven to fine tune the waveguide's phasematching condition for SHG at 1560.4 nm. The loss incurred through re-collimation of the 1560 nm pump using a B-coated lens is measured and used to calibrate the final measurements so that reported efficiencies are not artificially increased. The Thorlabs S121C silicon power meter is rated up to 500 mW. To validate its output at the higher powers measured in this work, SHG power values were verified against 3 dB and 10 dB filtering; this confirmed that the detector response used in this configuration remained linear at the powers measured. The dual-detector configuration enables the simultaneous measurement of both the launched pump power and SHG output. By measuring the pump power exiting the waveguide, the power that was launched into the crystal can be determined so that the crystal efficiency can be calculated by comparing with the power at the input facet; it should be noted that in some studies, particular reports of fiber coupled devices, the efficiency is calculated using power on the input facet, which creates a crystal efficiency dependent on launch efficiency.

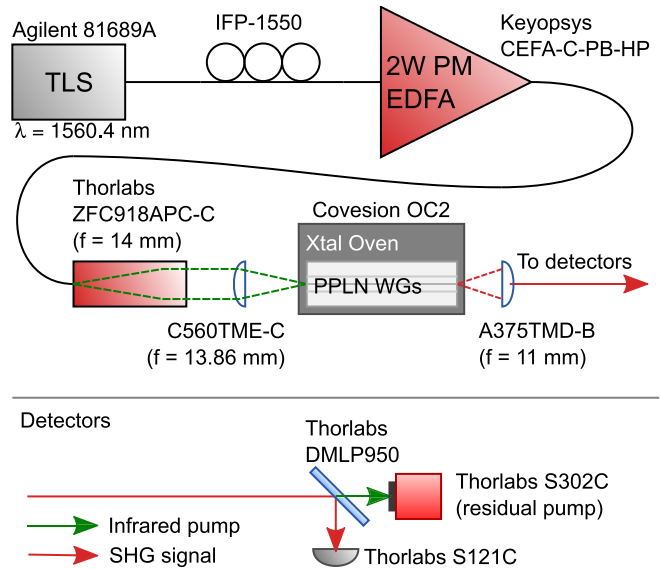

**Fig. 3.** Experimental setup used to characterise SHG conversion efficiency of our MgO:PPLN ridge waveguides. TLS: tunable laser source, IFP-1550: Inline fiber polarizer (1550 nm), ZFC: zoom fiber collimator, DMLP950: dichroic mirror (long pass 950 nm cut-on wavelength).

## 4. Waveguide characterisation

Measured SHG output power versus launched pump power is shown in Fig. 4 showing the quadratic scaling for the small pump depletion region and no evidence of SHG power rolloff due to parasitic nonlinear effects. The first 10 data points (up to 300 mW pump power), where the efficiency scales linearly from an undepleted pump, are fitted using linear regression to obtain a normalised conversion efficiency of 100.6 %/W (6.29 %/W cm$^2$), demonstrating that the high efficiencies at low power do not provide an appropriate metric for higher power operation. The apparent non-zero crossing of the efficiency slope is due to non-negligible conversion of background EDFA power at low input powers.

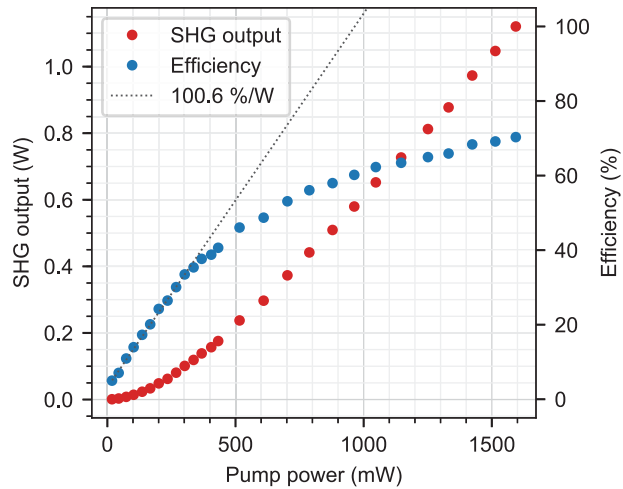

**Fig. 4.** SHG output power vs. launched pump power in the waveguide. At the maximum powers available in the lab, the PPLN waveguide did not exhibit signs of efficiency roll off due to parasitic nonlinear effects and the efficiency fit in the linear regime shows a normalised conversion efficiency of 100.6 %/W (6.29 %/W cm$^2$).

Figure 5 presents example SHG spectra from our waveguides at the maximum EDFA gain available in the lab where the efficiency did not decline over a 6 hour period in our initial measurements to ensure the EDFA, crystal temperature (113 °C), output power and spectrum were stable. 1.12 W of power was recorded at the target wavelength of 780.2 nm (the Rb$^{87}$ D$_2$ atomic cooling transition), with 1.59 W of 1560.4 nm pump measured at the waveguide output representing a crystal conversion efficiency of 70%. The inset shows the spectra when pumped with 160 mW of 1560 nm showing the spectra at low pump powers does not deviate significantly from its intrinsic sinc$^2$ profile across the powers used in this work; the enhancement of side lobes occur as the non-depleted pump approximation breaks down. The insertion loss is determined by comparing the 1560 nm power measured at the input facet at (2.08 W) with the power measured at the output facet away from the phasematching condition (1.59 W) at the maximum pump power, giving a total insertion loss of -1.17 dB. Low propagation losses necessitate anti-reflection coatings or angled facets due to the strong interface reflections causing large etalon fringes across the output spectra which are spectrally narrow and extremely sensitive to temperature changes in our 40 mm device lengths.

The waveguide output modes were measured on a calibrated NA setup using an unseeded EDFA source and a 780 nm SLED. The fundamental pump mode, shown top in Fig. 6, demonstrates near circular mode output with similar mode field diameters (MFDs) to single-mode telecoms fiber at 10.5 µm. Across the telecoms C-band the waveguide was single-mode, the ISO second-moment

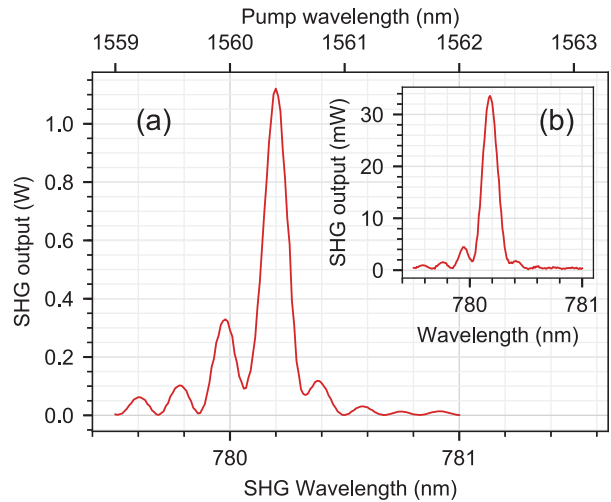

**Fig. 5.** (a) SHG spectra at highest launched pump power 1.59 W, measured exiting the waveguide. The crystal temperature was tuned to deliver the maximum output (113 °C) when pumped at 1560.4 nm to deliver maximum power at the Rb$^{87}$ D$_2$ line used for rubidium magneto-optical traps. The phasematching bandwidth is 0.3 nm FWHM, in agreement with the phasematching bandwidth expected for bulk PPLN. (b) The SHG spectra with 160 mW of pump power. The similarities in the spectra at these two extremes highlights its phasematching stability over the powers used in this study.

MFD was 9.91 µm and 11.14 µm for the *x* and *y* axes respectively (10.4 µm average MFD). When launched with the 780 nm SLED, the MFDs were measured at 11.22 µm and 8.99 µm for the *x* and *y* axes respectively. It is important to note that when launched with a 780 nm source, the waveguide is multimode whereas the phasematching condition of SHG ensures that only a single mode (the fundamental mode) is excited when used for frequency conversion. This means that the MFDs for 780 nm are indicative and a calibrated setup capable of measuring the SHG output directly would be required to measure the SHG single output mode.

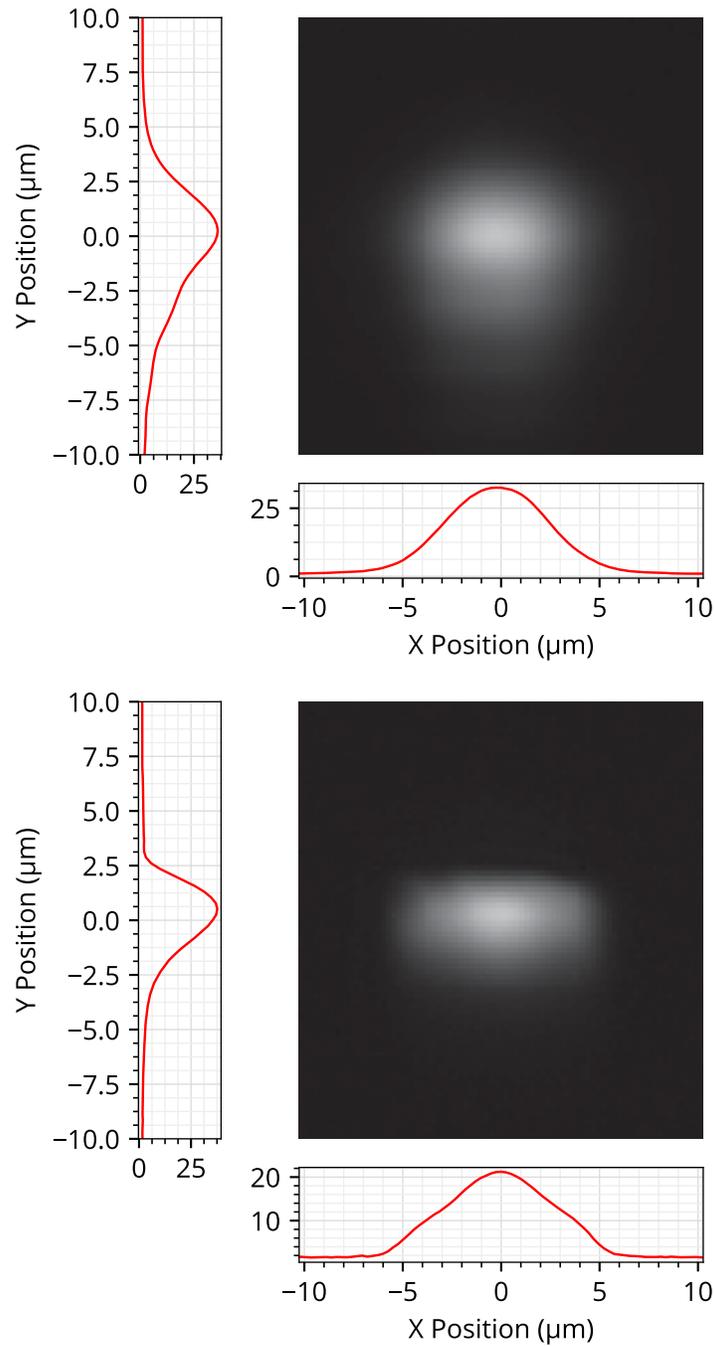

**Fig. 6.** Top - waveguide mode output at 1560 nm with 1-d averaged line plots demonstrating single mode output with near-circular ISO second-moment mode field diameter (MFD) of 9.91 µm and 11.14 µm for the *x* and *y* axes respectively. Bottom - waveguide mode output measured using a 780 nm SLED, the MFD is measured at 11.22 µm and 8.99 µm for the *x* and *y* axes respectively; however when launched with 780 nm light the waveguide, unlike its SHG output, is multimode.

## 5. Conclusion

In conclusion, we present the operation of our metallic zinc indiffused PPLN ridge waveguides at above 1.1 W of 780.2 nm SHG with 70% efficiency. This represents the highest reported value in the literature for this wavelength conversion regime using a commercially available C-band 2 W amplifier. The PPLN waveguide supports a near-circular mode with MFD 10.4 µm, which with AR coating enables efficient coupling of 1560 nm light with a total insertion loss of -1.17 dB.

## Funding

## Acknowledgements

The authors would like to acknowledge the support of Neil Sessions for assistance with cleanroom processing. Dr Corin Gawith acknowledges support from the Royal Academy of Engineering Senior Research Fellowship scheme.

## Disclosures

SAB+LGC: Covesion Ltd. (C,P), PGRS+CBEG: Covesion Ltd. (E,P). The authors declare no conflicts of interest.